\pdfoutput=1

\documentclass[11pt,twoside,a4paper,cmspaper,final,collab]{cms-tdr}
\def\svnVersion{9c08a68-D}\def\svnDate{2019/07/22}\def\cmsCernNoTag{CERN-EP-2018-288}\def\cmsCernDate{\today}\def\cmsMessage{"Published in Physics Letters B as \href{http://dx.doi.org/10.1016/j.physletb.2019.07.013}{\doi{10.1016/j.physletb.2019.07.013}.}"}

\begin{document}\cmsNoteHeader{HIG-18-003}

\hyphenation{had-ron-i-za-tion}
\hyphenation{cal-or-i-me-ter}
\hyphenation{de-vices}
\RCS$HeadURL$
\RCS$Id$

\newlength\cmsFigWidth
\ifthenelse{\boolean{cms@external}}{\setlength\cmsFigWidth{0.98\columnwidth}}{\setlength\cmsFigWidth{0.6\textwidth}}
\ifthenelse{\boolean{cms@external}}{\providecommand{\cmsLeft}{top}}{\providecommand{\cmsLeft}{left}}
\ifthenelse{\boolean{cms@external}}{\providecommand{\cmsRight}{bottom}}{\providecommand{\cmsRight}{right}}
\providecommand{\cmsTable}[1]{\resizebox{\textwidth}{!}{#1}}

\ifthenelse{\boolean{cms@external}}{\providecommand{\CL}{C.L.\xspace}}{\providecommand{\CL}{CL\xspace}}
\ifthenelse{\boolean{cms@external}}{\providecommand{\NA}{\ensuremath{\cdots}\xspace}}{\providecommand{\NA}{\ensuremath{\text{---}}\xspace}}
\newlength\cmsTabSkip\setlength{\cmsTabSkip}{1ex}

\cmsNoteHeader{HIG-18-003}

\newcommand{\gammaDark}{\ensuremath{{\PGg}_{\mathrm{D}}}\xspace}
\newcommand{\nOne}{\ensuremath{\textnormal{n}_{1}}\xspace}
\newcommand{\nDark}{\ensuremath{\textnormal{n}_{\mathrm{D}}}\xspace}
\newcommand{\PhOne}{\ensuremath{\Ph_{1}}\xspace}
\newcommand{\PhTwo}{\ensuremath{\Ph_{2}}\xspace}
\newcommand{\PhOneTwo}{\ensuremath{\Ph_{1,2}}\xspace}
\newcommand{\PaOne}{\ensuremath{\Pa_{1}}\xspace}
\newcommand{\PhI}{\ensuremath{\Ph_{i}}\xspace}
\newcommand{\dimuon}{\ensuremath{(\PGm\PGm)}\xspace}
\newcommand{\dimuonOne}{\ensuremath{(\PGm\PGm)_1}\xspace}
\newcommand{\dimuonTwo}{\ensuremath{(\PGm\PGm)_2}\xspace}

\newcommand{\MgammaDark}{\ensuremath{m_{{\PGg}_{\mathrm{D}}}}\xspace}
\newcommand{\MnOne}{\ensuremath{m_{{\textnormal{n}}_{1}}}\xspace}
\newcommand{\MnDark}{\ensuremath{m_{{\textnormal{n}}_{\mathrm{D}}}}\xspace}
\newcommand{\MPhOne}{\ensuremath{m_{\Ph_{1}}}\xspace}
\newcommand{\MPhTwo}{\ensuremath{m_{\Ph_{2}}}\xspace}
\newcommand{\MPhOneTwo}{\ensuremath{m_{\Ph_{1,2}}}\xspace}
\newcommand{\MPhI}{\ensuremath{m_{\Ph_{i}}}\xspace}
\newcommand{\MPa}{\ensuremath{m_{\Pa}}\xspace}
\newcommand{\MPaOne}{\ensuremath{m_{\Pa_{1}}}\xspace}
\newcommand{\Mdimuon}{\ensuremath{m_{(\PGm\PGm)}}\xspace}
\newcommand{\MdimuonOne}{\ensuremath{m_{(\PGm\PGm)_1}}\xspace}
\newcommand{\MdimuonTwo}{\ensuremath{m_{(\PGm\PGm)_2}}\xspace}
\newcommand{\Mmuon}{\ensuremath{m_{\PGm}}\xspace}
\newcommand{\Mtau}{\ensuremath{m_{\PGt}}\xspace}

\newcommand{\TgammaDark}{\ensuremath{\tau_{{\PGg}_{\mathrm{D}}}}\xspace}
\newcommand{\cTgammaDark}{\ensuremath{c\tau_{{\PGg}_{\mathrm{D}}}}\xspace}

\newcommand{\Zdimuon}{\ensuremath{z_{\dimuon}}\xspace}
\newcommand{\ZdimuonOne}{\ensuremath{z_{\dimuonOne}}\xspace}
\newcommand{\ZdimuonTwo}{\ensuremath{z_{\dimuonTwo}}\xspace}
\newcommand{\Ztrack}{\ensuremath{z_\text{track}}\xspace}
\newcommand{\Lxy}{\ensuremath{L_{xy}}\xspace}
\newcommand{\Lz}{\ensuremath{L_{z}}\xspace}
\newcommand{\DeltaY}{\ensuremath{\Delta y}\xspace}

\newcommand{\alphaGen}{\ensuremath{\alpha_\text{gen}}\xspace}
\newcommand{\epsilonFull}{\ensuremath{\epsilon_{\text{full}}}\xspace}
\newcommand{\SOne}{\ensuremath{S_\mathrm{I}(\Mdimuon)}\xspace}
\newcommand{\STwo}{\ensuremath{S_\mathrm{II}(\Mdimuon)}\xspace}
\newcommand{\SOneFinal}{\ensuremath{S_\mathrm{I}(\MdimuonOne)}\xspace}
\newcommand{\STwoFinal}{\ensuremath{S_\mathrm{II}(\MdimuonTwo)}\xspace}
\newcommand{\ASR}{\ensuremath{A_{\mathrm{SR}}}\xspace}
\newcommand{\ACR}{\ensuremath{A_{\mathrm{CR}}}\xspace}
\newcommand{\fSPS}{\ensuremath{f_\mathrm{SPS}}\xspace}

\newcommand{\muF}{\ensuremath{\mu_{\text{F}}}\xspace}
\newcommand{\muR}{\ensuremath{\mu_{\text{R}}}\xspace}

\newcommand{\functionMgammadark}{\ensuremath{f(\MgammaDark)}\xspace}

\newcommand{\PP}{\ensuremath{\Pp\Pp}\xspace}
\newcommand{\PhToTwoNOne}{\ensuremath{\Ph\to2\nOne}\xspace}
\newcommand{\PhOneTwoToPaOne}{\ensuremath{\PhOneTwo\to2\PaOne}\xspace}
\newcommand{\BrPhToGammaDarkX}{\ensuremath{\mathcal{B}(\Ph\to2\gammaDark+\PX)}\xspace}
\newcommand{\MassRangeNewLightBoson}{\ensuremath{0.25<\MPa<8.5\GeV}\xspace}
\newcommand{\NOneToDark}{\ensuremath{\nOne\to\nDark+\gammaDark}\xspace}
\newcommand{\EpsilonOverAlpha}{\ensuremath{\epsilonFull/\alphaGen}\xspace}
\newcommand{\BrPaOnetoPGm}{\ensuremath{\mathcal{B}(\PaOne\to2\PGm)}\xspace}
\newcommand{\PaOneToGluons}{\ensuremath{\PaOne\to\cPg\cPg}\xspace}

\newcommand{\SignalRegionBackgroundTotal}{\ensuremath{7.95\pm1.12\stat\pm1.45\syst}\xspace}
\newcommand{\ControlRegionbbbarEvents}{\ensuremath{43}\xspace}
\newcommand{\SignalRegionbbbarEvents}{\ensuremath{7.26\pm1.11\stat}\xspace}
\newcommand{\SignalRegionObservedEvents}{\ensuremath{9}\xspace}

\providecommand{\cmsTable}[1]{\resizebox{\textwidth}{!}{#1}}

\date{\today}

\abstract{
A search for new light bosons decaying into muon pairs is presented using a data sample corresponding to an integrated luminosity of 35.9\fbinv of proton-proton collisions at a center-of-mass energy $\sqrt{s} = 13\TeV$, collected with the CMS detector at the CERN LHC. The search is model independent, only requiring the pair production of a new light boson and its subsequent decay to a pair of muons. No significant deviation from the predicted background is observed. A model independent limit is set on the product of the production cross section times branching fraction to dimuons squared times acceptance as a function of new light boson mass. This limit varies between 0.15 and 0.39\unit{fb} over a range of new light boson masses from 0.25 to 8.5\GeV. It is then interpreted in the context of the next-to-minimal supersymmetric standard model and a dark supersymmetry model that allows for nonnegligible light boson lifetimes. In both cases, there is significant improvement over previously published limits.
}

\title{A search for pair production of new light bosons decaying into muons in proton-proton collisions at 13\TeV}

\hypersetup{%
pdfauthor={CMS Collaboration},%
pdftitle={A search for pair production of new light bosons decaying into muons in proton-proton collisions at 13 TeV},%
pdfsubject={CMS},%
pdfkeywords={CMS, New light boson, Supersymmetry, Hidden sector, Dark photon, Muon}}

\maketitle

\section{Introduction}
The standard model (SM) is known to give an incomplete description of particle physics and a number of extensions of the SM predict the existence of new light bosons \cite{Maniatis:2009re, AxionsDM, ArkaniHamed:2008qn}.
In this Letter, we present a model independent search for the pair production of
a light boson that decays into a pair of muons.
A simple example of pair production in proton-proton ($\PP$)~collisions is
${\Pp\Pp\to\Ph\to2\Pa+\PX\to4\PGm+\PX}$,
where \Ph is a Higgs boson (either SM or non-SM), \Pa is the new light neutral boson, and \PX are spectator particles that are predicted in several models~\cite{PhysRevD.81.075021}.
While production via the \Ph boson is possible, it is not required in the search presented here: the only requirement is that a pair of identical light bosons are created at a common vertex and each light boson subsequently decays to a pair of muons. These muon pairs are referred to as ``dimuons"; the dimuon and new light boson production vertices are allowed to be displaced.
The generic nature of this signature means that any limit set on the product of the cross section, branching fraction to dimuons squared, and acceptance is model independent; it can thus be reinterpreted in the context of specific models.

We develop a set of search criteria intended to minimize background events while remaining model independent. Two different classes of benchmark models are used to design the analysis and to verify that the results are actually model independent: the next-to-minimal supersymmetric standard model (NMSSM)~\cite{Fayet1975104, Kaul198236, Barbieri1982343, Nilles1983346, Frere198311, Derendinger1984307, Drees:1988fc, Maniatis:2009re, Ellwanger:2009dp} and supersymmetry (SUSY) models with hidden sectors (dark SUSY)~\cite{ArkaniHamed:2008qn, Baumgart:2009tn, Falkowski:2010cm}.
In the NMSSM benchmark models, two of the three charge parity (CP) even neutral Higgs bosons \PhOne or \PhTwo can decay to one of the two CP odd neutral Higgs bosons via $\PhOneTwoToPaOne$. The light boson \PaOne subsequently decays to a pair of oppositely charged muons; this is equivalent to \BrPaOnetoPGm.
In the dark SUSY benchmark models, the breaking of a new $U(1)_{\mathrm{D}}$ symmetry (where the subscript ``D" means ``Dark")  gives rise to a massive dark photon \gammaDark. This dark photon can couple to SM particles via a small kinetic mixing parameter $\varepsilon$ with SM photons. The lifetime, and thus the displacement, of the dark photon is dependent upon $\varepsilon$ and the mass of the dark photon \MgammaDark. The signal topologies investigated feature an SM-like Higgs boson \Ph that decays
via ${\PhToTwoNOne}$, where \nOne is the lightest non-dark neutralino. Both of the \nOne then decay via ${\NOneToDark}$, where \nDark is a dark neutralino that is undetected. The dark photon \gammaDark decays to a pair of oppositely charged muons.

This analysis contributes to an existing body of experimental work in the search for new light bosons. Previous searches at the LHC for ${\Ph\to2\Pa}$ include
4\PGm~\cite{Khachatryan:2015wka, ATLASmuJets13, Chatrchyan2011, PhysLettB.2013.09.009},
4\PGt~\cite{CMS-HIG-14-019},
4$\ell$~\cite{PhysRevD.92.092001, Aaboud:2018fvk},
4$\ell$/4\PGp~\cite{Aad:2014yea},
4$\ell$/8$\ell$~\cite{Aad:2015sms}, 
4\PQb~\cite{Aaboud:2016oyb, Aaboud:2018iil},
4\PGg~\cite{Aad:2015bua},
2\PQb2\PGt~\cite{CMS-HIG-17-024},
2\PGm2\PGt~\cite{Sirunyan:2018mbx},
and 6\PQq~\cite{Aaij2016} final states. A more thorough description of the NMSSM and dark SUSY models, their empirical and theoretical motivations, and constraints for their search set by previous experiments is included
in Refs.~\cite{Khachatryan:2015wka} and~\cite{PhysLettB.2013.09.009}.

The search presented in this Letter includes several improvements
compared to the previous results published by the CMS Collaboration on light boson pair production decaying to muons given in Ref.~\cite{Khachatryan:2015wka}.
The data used for this analysis correspond to an integrated luminosity of 35.9\fbinv of $\PP$ collisions at 13\TeV, compared to 20.7\fbinv at 8\TeV. While no dedicated analysis is performed targeting nonprompt decays, a new trigger with increased sensitivity to signatures with displaced vertices was implemented and the present search is also sensitive to signatures of this kind. The muon trigger uses reconstruction algorithms that do not rely on a primary vertex constraint for the track fit. In addition, no cut is applied on the displacement of the muon vertex with respect to the primary vertex.
Improvements were made to the CMS detector since Ref.~\cite{Khachatryan:2015wka}. Additional resistive plate chambers (RPCs) and cathode strip chambers (CSCs) in the outer layer of the CMS endcap muon system were installed along with improved readout electronics for the innermost CSCs. There is an upgraded hardware trigger that includes improved algorithms for the assignment of transverse momentum to muon candidates. There is also a new software trigger algorithm that uses three muons instead of two and does not require the muons to come from the interaction point. These changes are discussed in detail in Refs. \cite{CMS-MUO-16-001, Tapper:1556311}.
These changes have led to improved detection sensitivity and a greater coverage of model parameter space. 
The analysis criteria were modified to improve the detection sensitivity and allow greater coverage of model parameter space as compared to Ref.~\cite{Khachatryan:2015wka}. For the NMSSM benchmark models, this is a search for \PaOne with a mass between 0.25 and 3.55\GeV. For the benchmark dark SUSY models, this is a search for \gammaDark with a mass ranging from 0.25 to 8.5\GeV and lifetime up to $\cTgammaDark=100\mm$. The motivation for the values of these model parameters is given in Section \ref{sec:signal_modeling}.

\section{The CMS detector}
The central feature of the CMS apparatus is a superconducting solenoid of 6\unit{m} internal diameter, providing a magnetic field of 3.8\unit{T}. Within the solenoid volume are a silicon pixel and strip tracker, a lead tungstate crystal electromagnetic calorimeter, and a brass and scintillator hadron calorimeter, each composed of a barrel and two endcap sections.  Forward calorimeters extend the pseudorapidity ($\eta$) coverage provided by the barrel and endcap detectors. Muons are detected in gas-ionization chambers embedded in the steel flux-return yoke outside the solenoid.

Muons are measured in the range $\abs{\eta} < 2.4$, with detection planes made using three technologies: drift tubes, cathode strip chambers, and resistive plate chambers. For muons with $\pt>20\GeV$ the single muon trigger efficiency exceeds 90\% over the full $\eta$ range, and the efficiency to reconstruct and identify muons is greater than 96\%. Matching muons to tracks measured in the silicon tracker results in a relative transverse momentum resolution, for muons with \pt up to 100\GeV, of 1\% in the barrel and 3\% in the endcaps. The \pt resolution in the barrel is better than 7\% for muons with \pt up to 1\TeV~\cite{CMS-MUO-16-001}.

Events of interest are selected using a two-tiered trigger system~\cite{Khachatryan:2016bia}. The first level (L1), composed of custom hardware processors, uses information from the calorimeters and muon detectors to select events at a rate of around 100\unit{kHz} within a time interval of less than 4\mus. The second level, known as the high-level trigger (HLT), consists of a farm of processors running a version of the full event reconstruction software optimized for fast processing, and reduces the event rate below 1\unit{kHz} before data storage.

A more detailed description of the CMS detector, together with definitions of the coordinate system used and the relevant kinematic variables, can be found in Ref.~\cite{Chatrchyan:2008aa}.

\section{Data selection}
\label{sec:data_selection}

The data were collected with a trigger that uses muon reconstruction algorithms that have an efficiency greater than 80\% up to the maximum vertex displacement (98\mm) studied in this analysis~\cite{Abbiendi:2015txa}. This maximum vertex displacement is motivated in Section~\ref{sec:signal_modeling}. 
The HLT is seeded by requiring the presence of two muons selected by the L1 trigger in an event, the leading muon with $\pt>12\GeV$, the subleading muon with $\pt>5\GeV$, and both satisfying $\abs{\eta}<2.4$. 
Events that later pass the HLT are required to have at least three reconstructed muons: one with $\pt>15\GeV$ and $\abs{\eta}<2.4$, the other two with $\pt>5\GeV$ and $\abs{\eta}<2.4$. The final state particles in the events are reconstructed using the particle-flow (PF) algorithm which performs a global fit that combines information from each subdetector~\cite{Sirunyan:2017ulk}.

The offline event selection in this analysis requires events to have a primary vertex reconstructed using a Kalman filtering (KF) technique~\cite{Chatrchyan:2014fea}.
In addition, each event contains at least  four muons, reconstructed with the PF algorithm, and identified as muons either by the PF algorithm itself or by using additional information from the calorimeter and muon systems. Each muon is required to have $\pt>8\GeV$ and $\abs{\eta}<2.4$. At least one muon must be a ``high-\pt'' muon, \ie, it must be found in the barrel region ($\abs{\eta}<0.9$) and must have $\pt>17\GeV$ in order to ensure that the trigger reconstruction has high efficiency and has no dependence on $\eta$.

Dimuons are constructed from pairs of oppositely charged muons that share a common vertex, reconstructed using a KF technique, and must have an invariant mass $\Mdimuon$ less than 9\GeV. This restriction ensures that there is no contribution to the SM background from the \PZ boson decays nor the \PgU ~meson system. These muons pairs must not have any muons in common with one another. Exactly two dimuons must be present in each event. A dimuon that contains a high-\pt muon is called a ``high-\pt dimuon''.
When only one high-\pt muon is present in the event, the high-\pt dimuon is denoted as \dimuonOne, while the other is denoted as \dimuonTwo. When both dimuons have at least one high-\pt muon, the dimuons are labeled randomly to prevent a bias in kinematic distributions.
Single muons not included in dimuons are called ``orphan'' muons.
No requirement is applied on the number of orphan muons.
Each reconstructed dimuon must contain at least one muon that has at least one hit that is recorded by a layer of the pixel system. This requirement preserves the high reconstruction efficiency for our signal benchmark models.
The dimuons are required to originate from the same primary vertex, $\abs{\ZdimuonOne - \ZdimuonTwo}<0.1\cm$, where \Zdimuon is the $z$ position of the secondary vertex associated with the dimuon propagated back to the beamline along the dimuon direction vector.
Furthermore, each dimuon must be sufficiently isolated.
The dimuon isolation $I_{\dimuon}$ is calculated as the \pt sum of charged-particle tracks with $\pt>0.5\GeV$ in the vicinity of the dimuon within $\DR < 0.4$ and $\abs{\Ztrack - \Zdimuon}<0.1\cm$. Here, $\DR$ is defined in terms of the track separation in $\eta$ and azimuthal angle ($\phi$, in radians) as $\DR = \sqrt{\smash[b]{(\Delta\eta)^2+(\Delta\phi)^2}}$, while \Ztrack is defined as the $z$ coordinate of the point of closest approach to the primary vertex along the beam axis.
Tracks included in the dimuon reconstruction are excluded from the isolation calculation.
The total isolation sum must be less than 2\GeV. Since the dimuons are expected to originate from the same type of light bosons,
the dimuon masses should be consistent with each other to within five times the detector resolution.
This requirement carves out a signal region (SR) in the two-dimensional plane of the dimuon invariant masses \MdimuonOne and \MdimuonTwo. The signal region is illustrated in Fig.~\ref{fig:signal2D} (left).

\section{Signal modeling}
\label{sec:signal_modeling}

The $\PP$ collisions at $\sqrt{s}= 13\TeV$ are simulated for samples in each of the two benchmark models, NMSSM and dark SUSY.
The parton distribution functions (PDFs) are modeled using NNPDF2.3LO~\cite{Ball:2013hta}.
The underlying event activity at the LHC and jet fragmentation is modeled with the Monte Carlo (MC) event generator \PYTHIA~\cite{SJOSTRAND2015159} using the ``CUETP8M1'' tune~\cite{Khachatryan:2015pea}. Specifically, \PYTHIA 8.212 is used for NMSSM and \PYTHIA 8.205 for the dark SUSY models. In each model, only Higgs boson production via gluon-gluon (\cPg\cPg) fusion is considered.
A single mass point is also generated through vector boson fusion (VBF) and associated vector boson production (VH) to determine their contribution to the ${\PhTwo\to2\PaOne}$ rate; this is included in a simplified reference scenario discussed later.

In the case of the NMSSM,
a simulated Higgs boson, either \PhOne or \PhTwo (generically denoted by \PhOneTwo), is forced to decay to a pair of light bosons $\PaOne$.
Each \PaOne subsequently decays to a pair of oppositely charged muons.
Since the \PhOneTwo in $\PhOneTwoToPaOne$ might not be the observed SM Higgs boson \cite{Aad:2012tfa, Chatrchyan:2012xdj, Chatrchyan:2013lba},  mass values of \MPhOneTwo between 90 and 150\GeV are simulated. This range is motivated by constraints set by the relic density measurements from WMAP~\cite{Hinshaw:2012aka} and Planck~\cite{Ade:2013zuv}, as well as searches at LEP~\cite{Abbiendi:2002qp, Abbiendi:2002in, Schael:2006cr, Abbiendi:2004ww, Abdallah:2004wy, Schael:2010aw, Schael:2006cr}. The light boson mass is simulated to vary between 0.25 and 3.55\GeV, or approximately 2\Mmuon and 2\Mtau, as motivated in Ref. ~\citep{Dermisek:2010mg}.  

In the case of dark SUSY, production of SM Higgs bosons is simulated with the MC matrix-element generator \MADGRAPH 4.5.2~\cite{Alwall:2007st} at leading order.
The non-SM decay of the Higgs bosons is modeled using the {{\textsc{BRIDGE 2.24}}\xspace} program~\cite{Meade:2007js}. Higgs bosons are forced to decay to a pair of SUSY neutralinos \nOne via ${\PhToTwoNOne}$. Each SUSY neutralino in turn decays to a dark photon and a dark neutralino via {$\NOneToDark$}. The dark neutralino mass \MnDark is set to 1\GeV; they are considered stable and thus escape detection. We set the dark photons to decay to a pair of oppositely charged muons 100\% of the time, ${\gammaDark\to\PGmm\PGmp}$. Only signal events are generated because these MC generated events are used to determine the effect of the selection criteria on the signal. The Higgs boson and \nOne masses are fixed to 125 and 10\GeV, respectively.
Dark photon masses \MgammaDark are simulated between 0.25 and 8.5\GeV. The upper value was chosen such that any observed peak will be fully below the 9\GeV limit described in Section~\ref{sec:data_selection}. Since dark photons interact weakly with SM particles, their decay width is negligible compared to the resolution in the dimuon mass spectrum.  Muon displacement is modeled with an exponential distribution with $\cTgammaDark$ between 0 and 100\mm. All MC generated events are run through the full CMS simulation based on \GEANTfour~\cite{Agostinelli:2002hh} and reconstructed with the same algorithms that are used for data.  

One of the key features of this analysis is the model independence of the results. This is confirmed by verifying that the ratio of the full reconstruction efficiency \epsilonFull over the generator level acceptance \alphaGen is independent of the signal model. The signal acceptance is defined as the fraction of MC-generated events that pass the generator level selection criteria.
The criteria are as follows: at least four muons in each event with $\pt > 8\GeV$ and $\abs{\eta}<2.4$, at least one muon with $\pt > 17\GeV$ and $\abs{\eta}<0.9$, and both light bosons must have a transverse decay length $\Lxy<9.8\cm$ and longitudinal decay length $\abs{\Lz}<46.5\cm$. The upper limits on \Lxy and $\abs{\Lz}$ correspond to the dimensions of the outer layer of the CMS pixel system and define the volume in which a new light boson decay can be observed in this analysis..
The parameter \epsilonFull is defined as the fraction of MC-generated
events that pass the trigger and full offline selection described above.
The insensitivity to the model used is displayed in Table~\ref{tab:signal_acceptance}.

\begin{table*}[htb]
\centering
\topcaption{The full reconstruction efficiency over signal acceptance $\EpsilonOverAlpha$ in \% for several representative signal NMSSM (upper) and dark SUSY benchmark models (lower). All uncertainties are statistical.}
\begin{tabular}{l c c c c c}
\hline
$\MPhOne\,[\GeVns{}]$ & 90 & 100 & 110 & 125 & 150\\
$\MPaOne\,[\GeVns{}]$ & 2 & 0.5 & 3 & 1 & 0.75\\
\hline
$\epsilonFull\,[\%]$ &
$8.85 \pm 0.06$ & $13.23 \pm 0.08$ & $11.96 \pm 0.07$ & $14.68 \pm 0.08$ & $18.48 \pm 0.09$ \\
$\alphaGen\,[\%]$ & $13.93\pm 0.08$ & $20.47 \pm 0.09$ & $19.24 \pm 0.09$ & $23.59 \pm 0.10$ & $29.93 \pm 0.10$ \\
$\EpsilonOverAlpha\,[\%]$ & $63.52\pm 0.29$ & $64.62 \pm 0.24$ & $62.19 \pm 0.25$ & $62.23 \pm 0.22$ & $61.73 \pm 0.20$\\
\hline
\end{tabular}
\\[1cm]
\cmsTable{
\begin{tabular}{l ccc c ccc}
\hline
$\MgammaDark\,[\GeVns{}]$ & \multicolumn{3}{c}{$ 0.25 $} & \multicolumn{1}{c}{}& \multicolumn{3}{c}{$ 8.5 $} \\ \cline{2-4}\cline{6-8}
$\cTgammaDark$\,[mm] & 0 & 1 & 5 && 0 & 2 & 20\\
\hline
$\epsilonFull\,[\%]$ & $9.12 \pm 0.21$ & $1.72 \pm 0.06$ & $0.12\pm 0.01$ & & $12.78\pm 0.12$ & $12.25\pm 0.06$ & $3.61\pm 0.02$\\
$\alphaGen\,[\%]$ & $13.52 \pm 0.25$ & $2.85 \pm 0.07$ & $0.20\pm 0.01$ & &$20.49 \pm 0.14 $ & $20.05 \pm 0.08$ & $6.16 \pm 0.03$ \\
$\EpsilonOverAlpha\,[\%]$ & $67.47 \pm 0.91$ & $60.2 \pm 1.3$ & $58.39\pm 2.0$ && $62.36 \pm 0.38$ & $61.10 \pm 0.21$ & $58.70 \pm 0.24$ \\
\hline
\end{tabular}
}
\label{tab:signal_acceptance}
\end{table*}

Scale factors are determined to correct for the differences between observed data and simulated samples.
Corrections for the identification and isolation of muons and isolation of dimuons are measured using ${\PZ\to\PGmm\PGmp}$ and ${\JPsi\to\PGmm\PGmp}$ samples using a ``tag-and-probe'' technique \cite{Khachatryan:2010xn}; the samples used are events from simulated data and from observed data control regions enriched in events from the aforementioned SM processes.
All muons in these samples are required to have $\pt>8\GeV$, the ``tag'' muon is required to be a loose muon as described in Ref.~\cite{CMS-MUO-16-001}, while the ``probe'' muon criteria vary according to the variable under study.
Corrections for the trigger efficiency are calculated using ${\PW\PZ\to3\PGm}$ and ${\ttbar\PZ\to3\PGm}$ events in simulated samples and in control data samples enriched with those processes. The control data samples are selected using a missing transverse energy requirement such that the control data sample is primarily composed of events that are different from those in the data sample used in this analysis.

A scale factor per event obtained from the efficiency seen in data, $\epsilon_{\text{data}}$, compared to the efficiency seen in MC generated data, $\epsilon_{\text{sim}}$, is determined to be $\epsilon_{\text{data}}/\epsilon_{\text{sim}} = 0.93 \pm 0.06\stat$. 

\section{Background estimation}
\label{sec:bkg}
The selection criteria described in Section \ref{sec:data_selection} are effective at reducing and eliminating most SM backgrounds with similar topology to our signal.
As a result, this analysis is expected to have a very small background contribution in the SR.
Three SM backgrounds are found to be nonnegligible and are presented here: bottom quark pair production (\bbbar), prompt double \JPsi meson decays, and electroweak production of four muons. Contributions from \PgU ~mesons are also considered; they are found to be negligible below the 8.5\GeV upper bound on the mass of the new light boson. Cosmic ray backgrounds are negligible. The total background contribution in the SR is estimated to be \SignalRegionBackgroundTotal events; the  contributions from each process are described below.

\subsection{The \texorpdfstring{\bbbar}{b b-bar}background}
\label{subsec:bbbar}
The largest background, \bbbar production, is dominated by events in which both \PQb quarks decay
to $\PGmm\PGmp + \PX$ or decay through low-mass meson resonances such as $\omega$, \PGr, $\phi$, \JPsi, and \Pgy. The \JPsi meson decay contribution considered in this background is nonprompt; the prompt \JPsi meson decay contribution is discussed in Section~\ref{subsec:jpsi}. A minor contribution comes from events with charged particle tracks misidentified as muons.
A two-dimensional template $S(\MdimuonOne, \MdimuonTwo)$ is constructed in the plane of the two dimuon invariant masses and used to estimate the contribution to the SM background from \bbbar decays. The template is constructed as follows.

First, a \bbbar-enriched control sample is selected from events with similar kinematic properties as the signal events, but not included in the SR. Events are required to pass the signal trigger and have exactly three muons. One of these muons must have $\pt > 17\GeV$ within {$\abs{\eta}<0.9$}, while the other two have $\pt > 8\GeV$ within $\abs{\eta}<2.4$. In addition, the control sample selection requires a good primary vertex, exactly one dimuon, and one orphan muon. The longitudinal distance between the projections of the dimuon trajectory starting from its vertex and the orphan muon track back to the beam axis, $\Delta z ((\PGm\PGm), \PGm_\mathrm{orphan})$ must have an absolute value of less than $0.1\cm$. The dimuon is required to have at least one hit in the pixel system as explained in Section~\ref{sec:data_selection}. Finally, the dimuon isolation value cannot be higher than 2\GeV.

Next, two one-dimensional templates, \SOne and \STwo, are obtained from the \bbbar-enriched events. In the case of \SOne, at least one high-\pt muon is contained in the dimuon. In the case of \STwo, the high-\pt muon is the orphan muon and the dimuon may or may not contain another high-\pt muon. This procedure ensures that kinematic differences between signal events that have exactly two high-\pt dimuons or just one high-\pt dimuon are taken into account. Each distribution is fitted with a shape comprised of a Gaussian distribution for each light meson resonance, a double-sided Crystal Ball function~\cite{Oreglia:1980cs} for the \JPsi meson signal peak, and a set of sixth-degree Bernstein polynomials for the bulk background shape. The template $S(\MdimuonOne, \MdimuonTwo)$ is obtained as $\SOneFinal \otimes \STwoFinal$, where $\otimes$ represents the Cartesian product.

Finally, the two-dimensional template is normalized in the dimuon-dimuon mass space from 0.25 to 8.5 GeV. The template is represented as a function of  \MdimuonOne and \MdimuonTwo in Fig.~\ref{fig:signal2D} (left) by a gray scale. The SR defined in Section~\ref{sec:data_selection} is outlined by dashed lines. The region of the mass space outside the SR represent the control region for the \bbbar background. The ratio between the integral of the template in the SR \ASR and the control region \ACR is calculated to be $\text{R}=\ASR/\ACR=0.1444/0.8556$. The same figure also shows the \ControlRegionbbbarEvents events found in the data that pass all selection criteria except for the $\MdimuonOne \simeq \MdimuonTwo$ requirement and thus fall outside the SR. The number of \bbbar events in the SR is then estimated to be $(\ControlRegionbbbarEvents \pm \sqrt{\ControlRegionbbbarEvents})\,\text{R} = \SignalRegionbbbarEvents$.

This method of estimating the \bbbar contribution to background events is further validated by repeating the procedure for different dimuon isolation values (5, 10, 50\GeV) and without any isolation. The \bbbar event yield is stable in the SR within 20\%, which is assigned as a systematic uncertainty.

\subsection{Prompt double \texorpdfstring{\JPsi}{J/Psi} meson background}
\label{subsec:jpsi}
Two mechanisms contribute to prompt double \JPsi meson production: single parton scattering (SPS) and double parton scattering (DPS); these processes have been measured by CMS and ATLAS \cite{Khachatryan:2014iia, Aaboud:2016fzt}. They can mimic the signal process when each \JPsi meson decays to a pair of muons.
The prompt double \JPsi meson decay background is estimated with a method that uses both experimental and simulated data.
In a control sample of experimental data, the prompt and nonprompt double \JPsi meson decay contributions are separated using the matrix method (also called the ``ABCD" method~\cite{ABCD_method}). The prompt contribution is then extrapolated into the SR.
Double \JPsi meson events are selected with a trigger dedicated to bottom quark physics. Each event is required to have at least four muons with $\pt>3.5\GeV$ within $\abs{\eta}<2.4$. No high-\pt muon is required. Events must have exactly two dimuons, with labels \dimuonOne or \dimuonTwo assigned randomly. The dimuon isolation follows the same definition as in Section~\ref{sec:data_selection}.
The kinematic properties of SPS and DPS events are studied using MC simulation. These events are generated using \PYTHIA 8.212 and \HERWIG 2.7.1~\cite{Bahr:2008pv}.
The variable with the best SPS--DPS separation power is found to be the absolute difference in rapidity between the two dimuons, $\abs{\DeltaY}$.
To remove nonresonant muon pairs from the sample, the dimuon masses are required to be within 2.8 and 3.3\GeV. The ABCD method is then employed using the dimuon isolation values as uncorrelated variables in the plane $(I_{\dimuonOne}, I_{\dimuonTwo})$. The maximum isolation on \dimuonOne and \dimuonTwo is set to 12\GeV. Here, region ``A" is the region bounded by $I_{(\PGm\PGm)_{1,2}}<2\GeV$. Conversely, ``B", ``C", and ``D" are nonisolated sideband regions used to extrapolate the nonprompt contribution into region ``A".
The nonprompt $\abs{\DeltaY}$ distribution is determined from the sideband regions; this distribution is scaled to match the nonprompt contribution in region ``A". This is then subtracted from the $\abs{\DeltaY}$ distribution, leaving the prompt $\abs{\DeltaY}$ distribution in region ``A".
To separate the prompt SPS from prompt DPS in data, a template distribution $\fSPS\abs{\DeltaY_\text{SPS}}+(1-\fSPS)\abs{\DeltaY_\text{DPS}}$ is fitted to the corresponding $\abs{\DeltaY}$ distribution in data, where $\fSPS$ and $1-\fSPS$ are the fractions of prompt SPS and DPS events, respectively. Finally, this result is used to determine the number of events that are expected in the SR of our experimental data sample.
The contribution of the prompt double \JPsi meson decay events in data passing the signal selections in Section~\ref{sec:data_selection} is calculated to be $N_\text{data}(\mathrm{SR}) = 0.33 \pm 0.08\stat \pm 0.05 \syst$.

\subsection{Electroweak background}
\label{subsec:EWK}
Electroweak production of four muons, {$\Pp\Pp\to4\PGm$}, is estimated using MC events generated with \CALCHEP 3.6.25~\cite{Belyaev:2012qa}. The processes studied  include ${\qqbar\to \PZ\PZ^* \to 2\PGmm2\PGmp}$ and  $\qqbar\to\PZ\to\PGmm\PGmp$, where one of the muons radiates a second \PZ boson that decays to a $\PGmm\PGmp$ pair. Other electroweak processes, such as ${\Pp\Pp\to\Ph(125)\to \PZ\PZ^*\to2\PGmm2\PGmp}$,  are determined to be negligible a priori and thus are not included. Based on the simulation, the electroweak background is found to be $0.36\pm0.09\stat$. Unlike the prompt double \JPsi meson decay background, the electroweak background is not concentrated at any particular mass value; its contribution to any mass bin is negligible compared to the \bbbar background. Consequently, these background events are neglected in any limit setting computation.

\begin{figure*}[htb]
\centering
\includegraphics[width=0.50\linewidth]{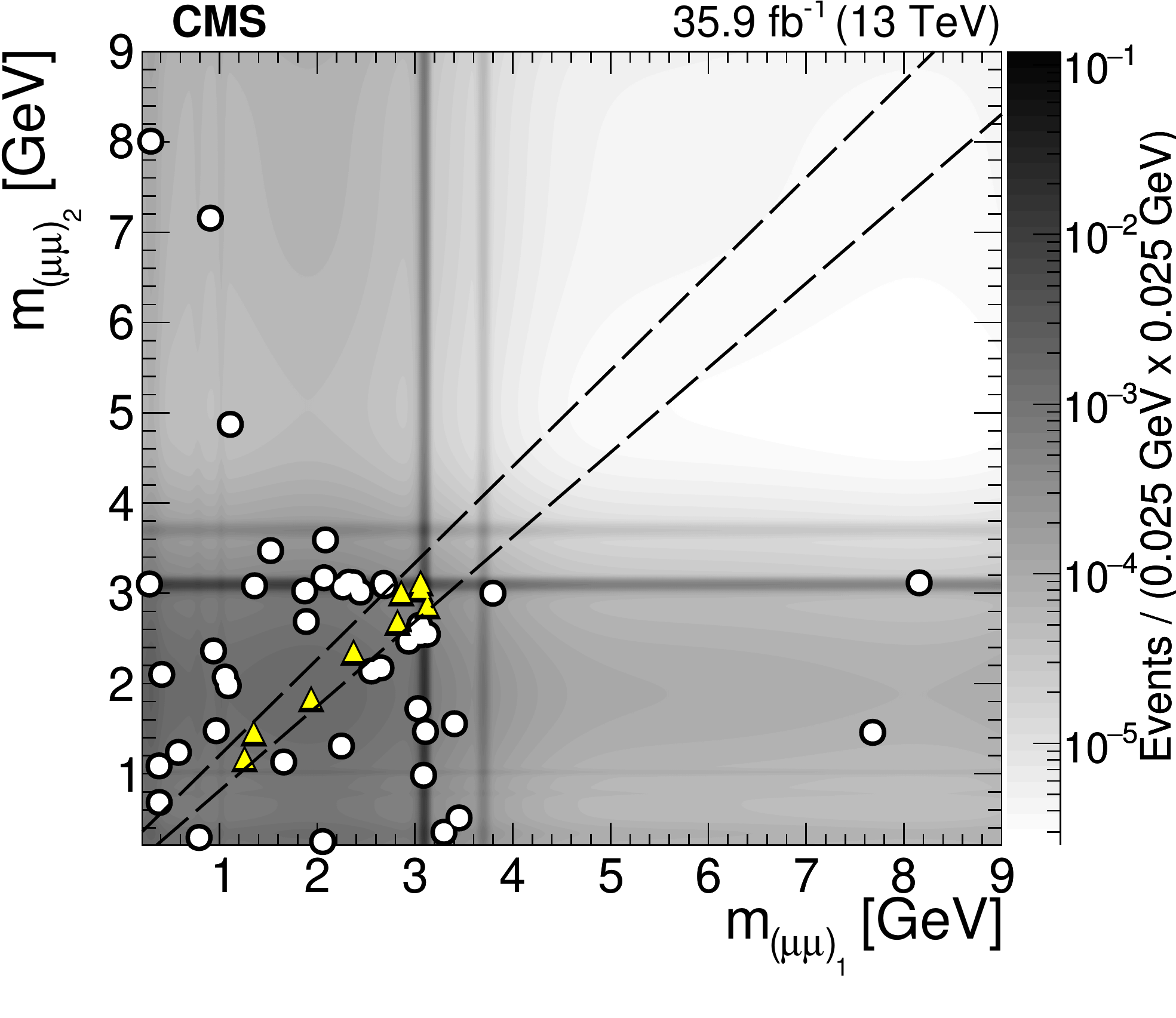}
\hfill
\includegraphics[width=0.45\linewidth]{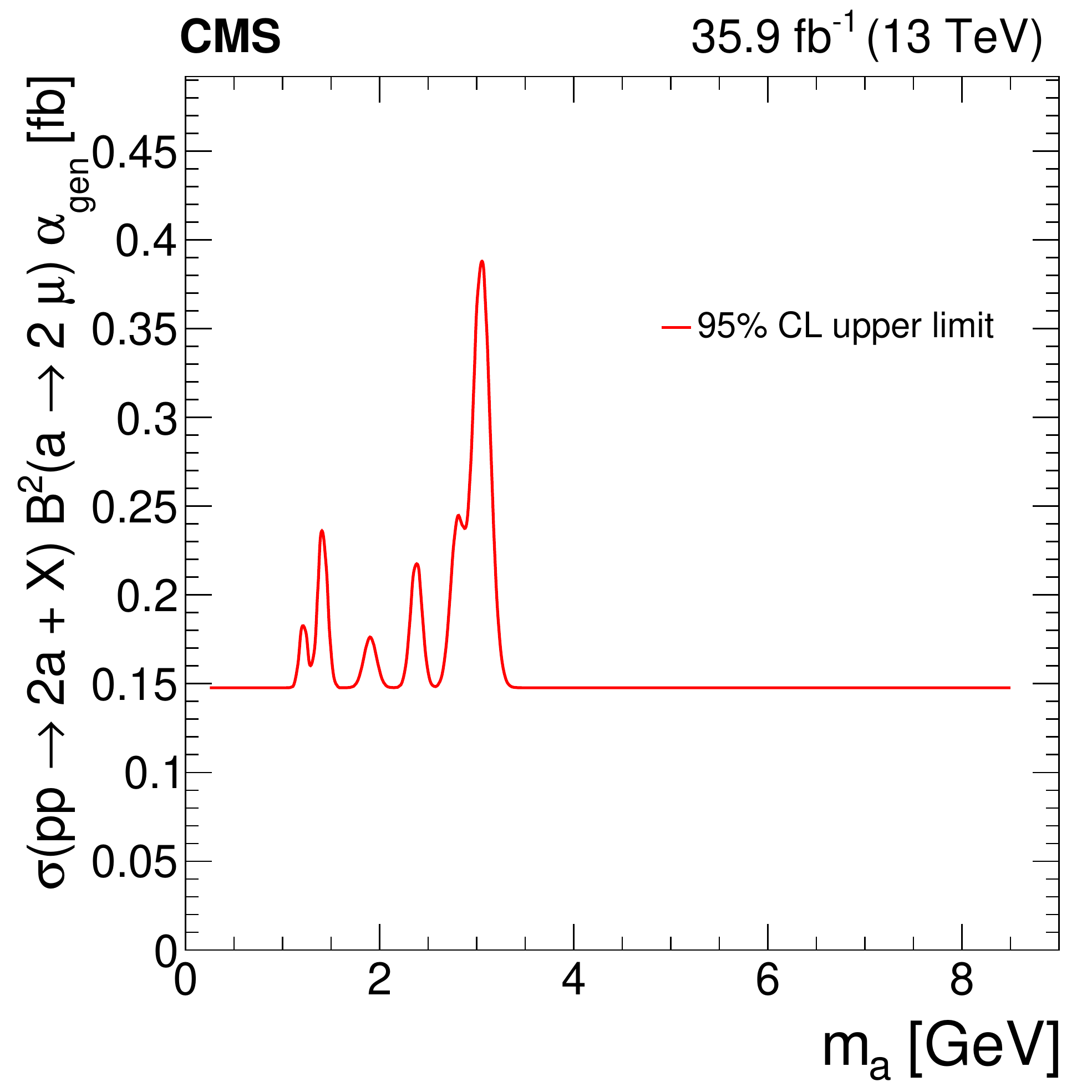}
\caption{Left: Distribution of the invariant masses \MdimuonOne vs. \MdimuonTwo of the isolated dimuon systems; triangles represent data events passing all the selection criteria and falling in the SR $\MdimuonOne \simeq \MdimuonTwo$ (outlined by dashed lines); white bullets represent data events that pass all selection criteria but fall outside the SR. The grayscale heatmap shows the normalized distribution of expected events in the \bbbar background template. 
Right: The 95\% \CL upper limit set on ${\sigma(\Pp\Pp\to2\Pa+\PX)\mathcal{B}^2(\Pa\to2\PGm)\alphaGen}$ over the range ${\MassRangeNewLightBoson}$.}
\label{fig:signal2D_ModelIndependentLimit}
\label{fig:signal2D}
\label{fig:ModelIndependentLimit}
\end{figure*}

\section{Systematic uncertainties}
\label{sec:syst}

Both instrumental and theoretical sources of uncertainty are considered in this section. The leading source of instrumental uncertainty is the triple-muon trigger scale factor (6\%). It is dominated by the statistical uncertainty in events in the control region used to measure the scale factor. Other sources of instrumental uncertainty include the uncertainty in the measurement of the integrated luminosity recorded by the CMS detector (2.5\%)~\cite{CMS-PAS-LUM-17-001}, the muon identification data-to-simulation scale factor (0.6\% per muon for all simulated muons), the reconstruction of the dimuon in the tracker (1.2\% per dimuon) and in the muon system (1.3\% per dimuon) from spatially close muons, and the effect on the acceptance of the dimuon mass shape used to determine the width of  the SR (1.5\%). The uncertainty in the dimuon isolation and the contributions of extraneous $\PP$~collisions are determined to be negligible.

The theoretical uncertainties are dominated by the uncertainty in the PDFs, knowledge of the strong coupling constant \alpS, and the renormalization (\muR) and factorization (\muF) scales. The PDF and \alpS uncertainties are estimated using a technique that follows the PDF4LHC recommendations~\cite{PDF4LHC_RUNII, Ball:2012cx}. The uncertainty in the scale factors is determined by simultaneously varying \muR and \muF up and down by a factor of two using \textsc{MCFM} 8.0~\cite{CAMPBELL201010}.
The effect of PDF choice and PDF parameter variation upon the central values is also studied. When all previously described theoretical uncertainties are added in quadrature, the sum is 8\%.
The uncertainty in the branching fraction ${\mathcal{B}(\Ph\to2\Pa+\PX\to4\PGm+\PX)}$ is taken to be 2\%~\cite{Chatrchyan:2013lba}.

\section{Results}
\label{sec:results}

\begin{figure*}[tbh]
\centering
\includegraphics[width=0.49\linewidth]{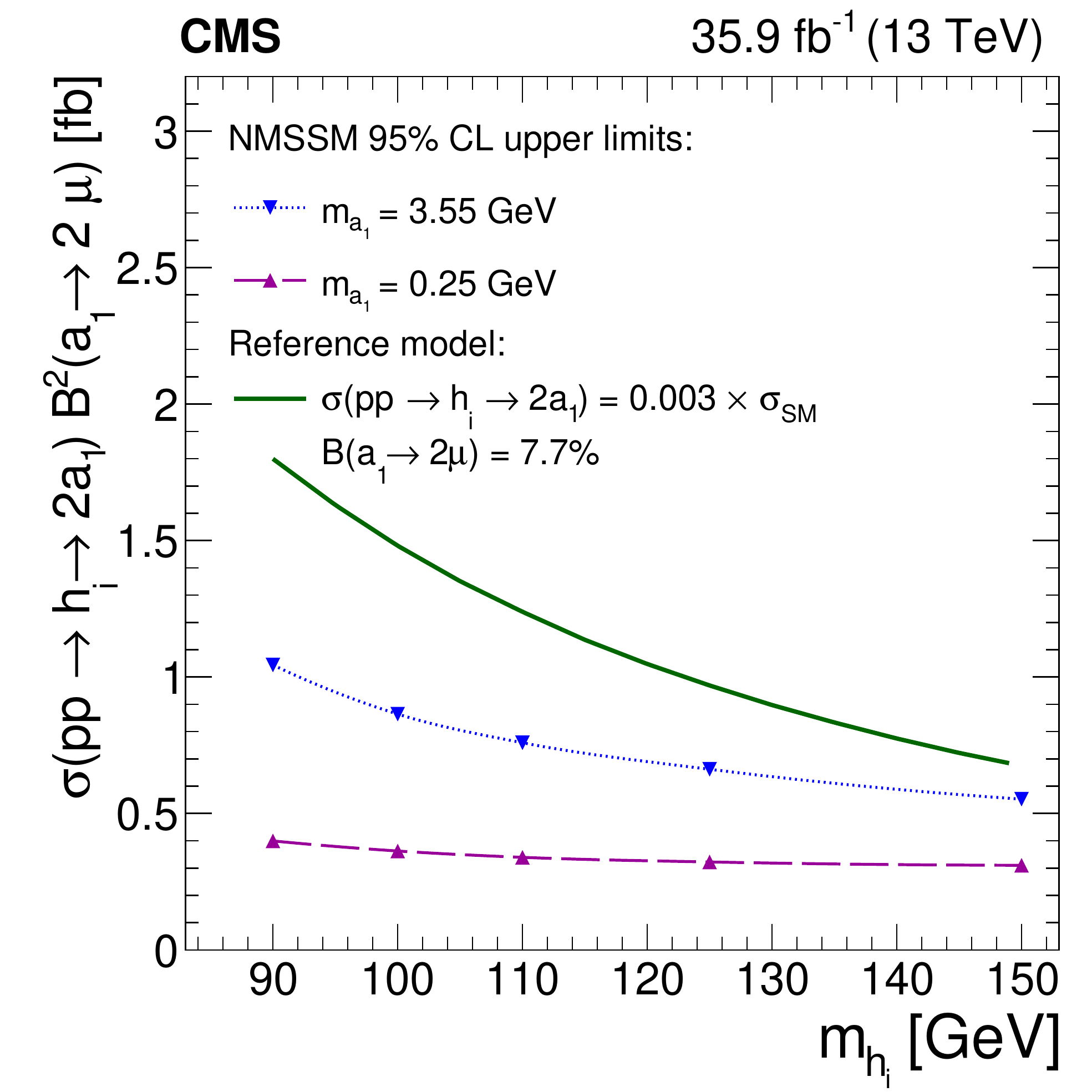}
\hfill
\includegraphics[width=0.49\linewidth]{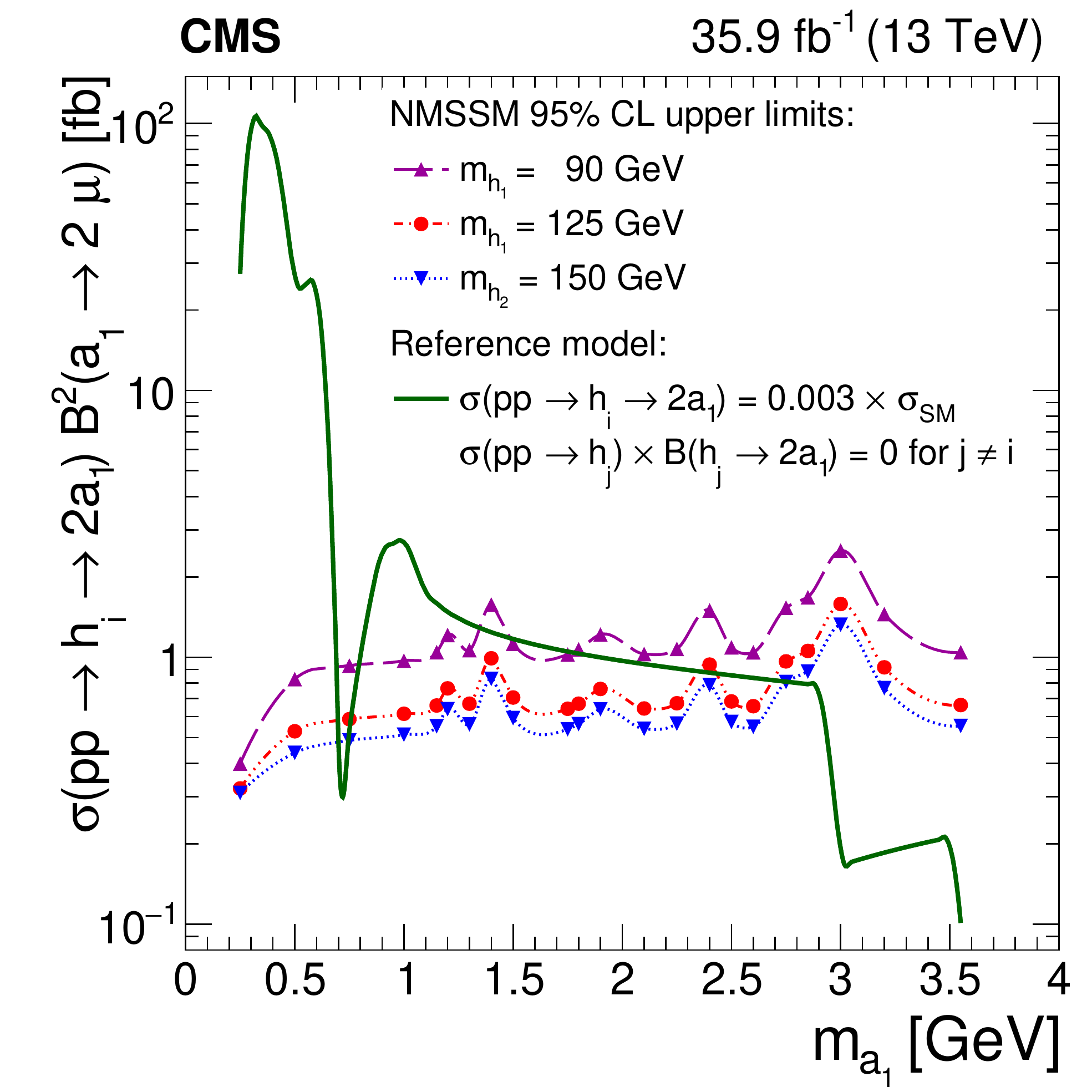}
\caption{Left: The limits are compared to a representative model (solid curve) obtained using the simplified scenario described in the text. The figure is separated into two regions: $\MPhI=\MPhOne < 125\GeV$ with $\MPhTwo = 125\GeV$, and $\MPhOne = 125\GeV$ with $\MPhI=\MPhTwo  > 125\GeV$.
Right: These limits are compared to a representative model (solid curve) from the simplified scenario described in the text. The simplified scenario includes \cPg\cPg-fusion, VBF, and VH production modes.
}
\label{fig:limits_vs_mh}
\end{figure*}

After applying all selection criteria to the data sample, \SignalRegionObservedEvents events are found in the SR. Their distribution in  \MdimuonOne and \MdimuonTwo is shown in Fig.~\ref{fig:signal2D} (left). This result is consistent with the sum of all background estimates described in Section~\ref{sec:bkg}, which is found to be \SignalRegionBackgroundTotal events.
A model independent 95\% confidence level (\CL) upper limit is set on the product of the production cross section times branching fraction to dimuons squared times acceptance. Limits are set using the \CLs method~\cite{CLS1,Junk:1999kv}. The test statistic used is based on the logarithm of the likelihood ratio~\cite{Khachatryan:2014jba}. The systematic uncertainties and their correlations have been
accounted for by profiling the likelihood with respect to the nuisance parameters for each value of the signal strength s; this results in the profile likelihood being a function only of s.
The limit is shown as a function of \MPa in Fig.~\ref{fig:signal2D} (right) over the range $\MassRangeNewLightBoson$; the limit varies between 0.15 and 0.39\unit{fb}. Neglecting the large peak in the upper limit at the \JPsi meson mass, the largest upper limit is 0.25\unit{fb}.
This result can be interpreted in the context of specific models.

For the NMSSM scenario, the 95\% \CL upper limit is derived for
$\sigma\left( \Pp\Pp\to\PhOneTwo\to2\PaOne\right)\mathcal{B}^2(\PaOne\to2\PGm)$
as a function of \MPhOneTwo for two choices of \MPaOne as shown in Fig.~\ref{fig:limits_vs_mh}~(left)
and as a function of \MPaOne for three choices of \MPhOne as shown in Fig.~\ref{fig:limits_vs_mh}~(right).
Since the choice of \MPhOne does not restrict \MPhTwo, we choose to set $\epsilonFull(\MPhTwo)=\epsilonFull(\MPhOne)$ to simplify the expression.
This choice is conservative because $\epsilonFull(\MPhTwo)>\epsilonFull(\MPhOne)$ if $\MPhTwo>\MPhOne$, for any \MPhOne.
In this simplified scenario, $\BrPaOnetoPGm$ is a function of \MPhOne as calculated in Ref.~\cite{Dermisek:2010mg}.
To facilitate comparison between the upper limits derived from this analysis and upper limits following from setting parameters in theoretical models, we include reference curves (solid line) in both Fig.~\ref{fig:limits_vs_mh} left and right.
For both reference curves, the ratio of the vacuum expectation values of the Higgs doublets \tanb is set to 20. We also set $\sigma(\Pp\Pp\to\PhI)=\sigma_\mathrm{SM}(\MPhI)$~\cite{Dittmaier:2011ti} and $\mathcal{B}(\MPhI \to2\PaOne)=0.3\%$ so that the resulting reference curves are similar to the upper limits that are determined from the yield of dimuon pair events observed in the data.
In Fig.~\ref{fig:limits_vs_mh}~(left), the reference curve is constructed with the assumption that $\BrPaOnetoPGm=7.7\%$ and $\MPaOne\approx2\GeV$. In the region where $\MPhI<125\GeV$, $\MPhOne$ is the independent variable and it is assumed that $\MPhTwo$ is the mass of the observed 125\GeV Higgs boson. In the region where $\MPhI > 125\GeV$, $\MPhTwo$ is the independent variable and it is assumed that $\MPhOne$ is the observed Higgs boson mass. Compared to the upper limits shown in Refs.~\cite{Khachatryan:2015wka}, Fig.~\ref{fig:limits_vs_mh}~(left) represents an improvement of a factor of {$\approx$1.5} for {$\MPaOne =3.55\GeV$} (dotted curve) and a factor of {$\approx$3} for {$\MPaOne=0.25\GeV$} (dashed curve).
In Fig.~\ref{fig:limits_vs_mh}~(right), we present $95\%$ \CL upper limits as functions of \MPaOne in the NMSSM scenario on $\sigma(\Pp\Pp\to\PhI\to2\PaOne)\mathcal{B}^2(\PaOne\to2\PGm)$ with $\MPhOne=90\GeV$ (dashed curve), $\MPhOne=125\GeV$ (dash-dotted curve), and $\MPhTwo=150\GeV$ (dotted curve). It is assumed that all contributions come from either \PhOne or \PhTwo; there is no case in which both \PhOne and \PhTwo decay to the \PaOne.
The sharp inflections in the reference curve are due to the fact that $\BrPaOnetoPGm$ is affected by the $\PaOne\to\cPqs\cPaqs$ and $\PaOneToGluons$ channels~\cite{Dermisek:2010mg}. As \MPhOne crosses the internal quark loop thresholds, $\mathcal{B}(\PaOneToGluons)$ changes rapidly, giving rise to structures in $\BrPaOnetoPGm$ at these values of \MPhOne.

\begin{figure}[tbh]
\centering
\includegraphics[width=\cmsFigWidth]{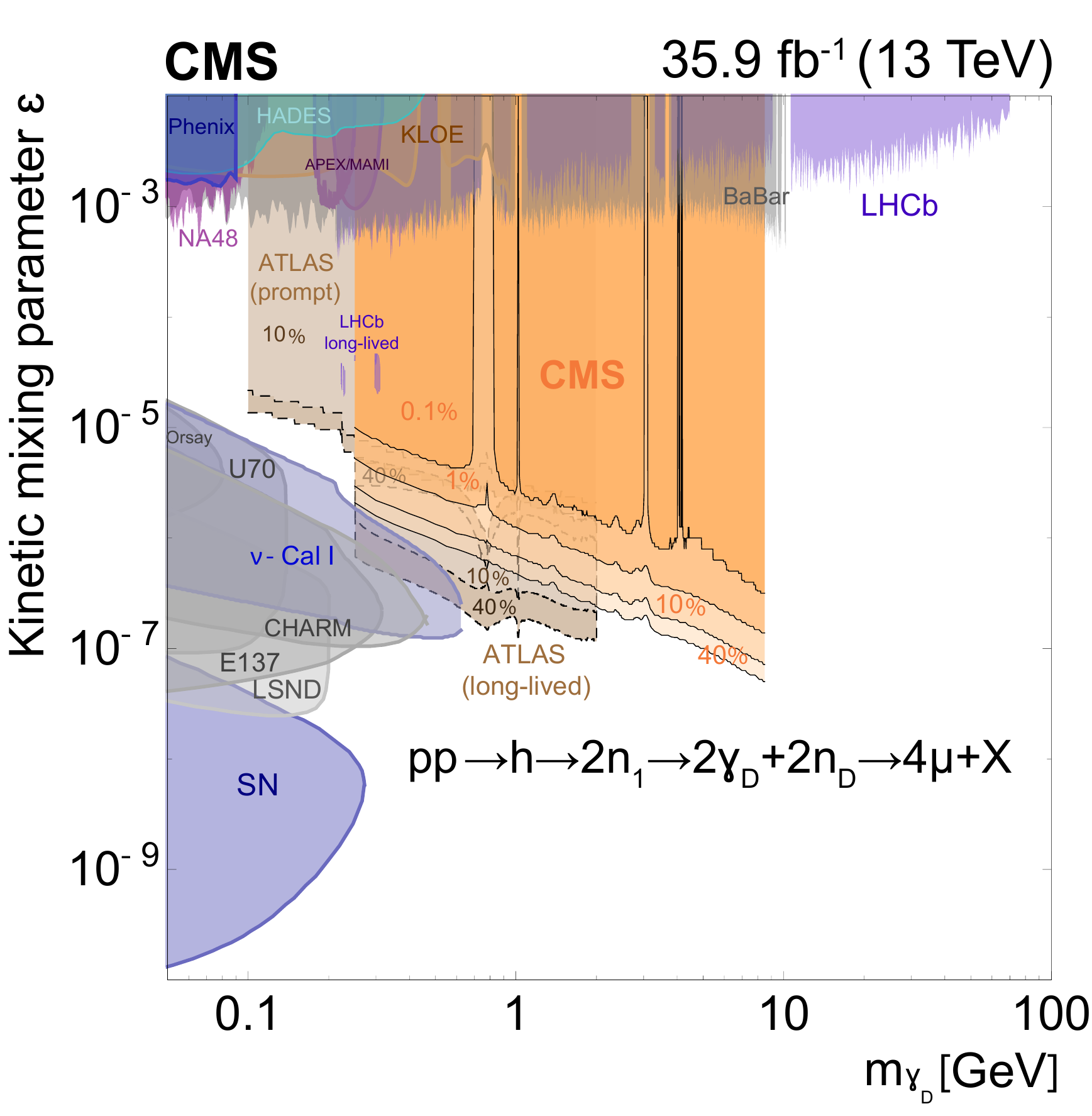}
\caption{
The 90\% \CL upper limits (black solid curves) from this search as interpreted in the dark SUSY scenario, where the process is
$\Pp\Pp\to\Ph\to2\nOne\to2\gammaDark+2\nDark\to4\PGm+\PX$, with $\MnOne=10\GeV$, and $\MnDark=1\GeV$.
The limits are presented in the plane of the parameters ($\varepsilon$ and \MgammaDark).
Constraints from other experiments~\cite{Aad:2014yea, Adare:2014mgk, Babusci:2014sta, Abrahamyan:2011gv, Merkel:2014avp, Agakishiev:2013fwl, Lees:2014xha, Fradette:2014sza,Essig:2013lka,2012arXiv1201.2683D,Dreiner:2013mua,Blumlein:2011mv,Essig:2010gu,Batell:2009di,Gninenko:2012eq, PhysRevLett.120.061801, Aad:2015sms} showing their 90\% \CL exclusion contours are also presented. The colored contours for the CMS and ATLAS limits represent different values of $\BrPhToGammaDarkX$ that range from 0.1 to 40\%.}
\label{fig:currentDarkSusyCostraints}
\end{figure}

For the dark SUSY scenario, a 90\% \CL upper limit is set on the product of the Higgs boson production cross section and the branching fractions of the Higgs boson (cascade) decay to a pair of dark photons.
The limit set by this experimental search is presented in Fig.~\ref{fig:currentDarkSusyCostraints} as areas excluded in a two-dimensional plane of $\varepsilon$ and \MgammaDark. Also included in Fig.~\ref{fig:currentDarkSusyCostraints} are limits from other experimental searches~\cite{Aad:2014yea, Adare:2014mgk, Babusci:2014sta, Abrahamyan:2011gv, Merkel:2014avp, Agakishiev:2013fwl, Lees:2014xha, Fradette:2014sza,Essig:2013lka,2012arXiv1201.2683D,Dreiner:2013mua,Blumlein:2011mv,Essig:2010gu,Batell:2009di,Gninenko:2012eq, PhysRevLett.120.061801, Aad:2015sms}. For both this search and the ATLAS searches, limits are shown for values of $\BrPhToGammaDarkX$ in the range 0.1--40\%. It should be noted that the 40\% value is excluded by the latest results on the branching fraction of the Higgs boson decay to invisible particles~\cite{Khachatryan:2017inv}. It serves merely for a comparison with limits obtained in a previous version of this search~\cite{Khachatryan:2015wka}. The kinetic mixing parameter $\varepsilon$, the mass of the dark photon \MgammaDark, and the lifetime of the dark photon \TgammaDark are related via an analytic function \functionMgammadark that is solely dependent on the dark photon mass~\cite{Batell:2009yf}; namely, $\TgammaDark(\varepsilon, \MgammaDark) = \varepsilon^{-2}\functionMgammadark$. The lifetime of the dark photon is allowed to vary from 0 to 100\mm and \MgammaDark can range from 0.25 to 8.5\GeV. Because of the extensions in the ranges of these parameters, this search constrains a large and previously unexplored area in the $\varepsilon$ and \MgammaDark parameter space. The limits on $\varepsilon$ presented in this Letter improve on those in Ref.~\cite{Khachatryan:2015wka} by a factor of approximately 2.5.

\section{Summary}
A search for pairs of new light bosons that subsequently decay to pairs of oppositely charged muons is presented. This search is developed in the context of a Higgs boson decay, {$\Ph\to2\Pa+\PX\to4\PGm+\PX$} and is performed on a data sample collected by the Compact Muon Solenoid experiment in 2016 that corresponds to an integrated luminosity of 35.9\fbinv proton-proton collisions at 13\TeV.
This data set is larger and collected at a higher center-of-mass energy than the previous CMS search~\cite{Khachatryan:2015wka}.
Additionally, both the mass range of the light boson \Pa and the maximum possible displacement of its decay vertex are extended compared to the previous version of this analysis.
Nine events are observed in the signal region (SR), with \SignalRegionBackgroundTotal events expected from the standard model (SM) backgrounds.
The distribution of events in the SR is consistent with SM expectations. A model independent 95\% confidence level upper limit on the product of the production cross section times branching fraction to dimuons squared times acceptance is set over the mass range ${\MassRangeNewLightBoson}$ and is found to vary between 0.15 and 0.39\unit{fb}.
This model independent limit is then interpreted in the context of dark supersymmetry (dark SUSY) with nonnegligible light boson lifetimes of up to $\cTgammaDark=100\mm$ and in the context of the next-to-minimal supersymmetric standard model (NMSSM). For the dark SUSY interpretation, the upper bound of \MgammaDark was increased from 2 to 8.5\GeV and the excluded $\varepsilon$ was improved by a factor of approximately 2.5. In the NMSSM, the 95\% \CL upper limit was improved by a factor of {$\approx$1.5\,(3)} for {$\MPaOne =3.55\,(0.25)$}\GeV over previously published limits.

\begin{acknowledgments}

We congratulate our colleagues in the CERN accelerator departments for the excellent performance of the LHC and thank the technical and administrative staffs at CERN and at other CMS institutes for their contributions to the success of the CMS effort. In addition, we gratefully acknowledge the computing centers and personnel of the Worldwide LHC Computing Grid for delivering so effectively the computing infrastructure essential to our analyses. Finally, we acknowledge the enduring support for the construction and operation of the LHC and the CMS detector provided by the following funding agencies: BMBWF and FWF (Austria); FNRS and FWO (Belgium); CNPq, CAPES, FAPERJ, FAPERGS, and FAPESP (Brazil); MES (Bulgaria); CERN; CAS, MoST, and NSFC (China); COLCIENCIAS (Colombia); MSES and CSF (Croatia); RPF (Cyprus); SENESCYT (Ecuador); MoER, ERC IUT, and ERDF (Estonia); Academy of Finland, MEC, and HIP (Finland); CEA and CNRS/IN2P3 (France); BMBF, DFG, and HGF (Germany); GSRT (Greece); NKFIA (Hungary); DAE and DST (India); IPM (Iran); SFI (Ireland); INFN (Italy); MSIP and NRF (Republic of Korea); MES (Latvia); LAS (Lithuania); MOE and UM (Malaysia); BUAP, CINVESTAV, CONACYT, LNS, SEP, and UASLP-FAI (Mexico); MOS (Montenegro); MBIE (New Zealand); PAEC (Pakistan); MSHE and NSC (Poland); FCT (Portugal); JINR (Dubna); MON, RosAtom, RAS, RFBR, and NRC KI (Russia); MESTD (Serbia); SEIDI, CPAN, PCTI, and FEDER (Spain); MOSTR (Sri Lanka); Swiss Funding Agencies (Switzerland); MST (Taipei); ThEPCenter, IPST, STAR, and NSTDA (Thailand); TUBITAK and TAEK (Turkey); NASU and SFFR (Ukraine); STFC (United Kingdom); DOE and NSF (USA).

\hyphenation{Rachada-pisek} Individuals have received support from the Marie-Curie program and the European Research Council and Horizon 2020 Grant, contract No. 675440 (European Union); the Leventis Foundation; the A. P. Sloan Foundation; the Alexander von Humboldt Foundation; the Belgian Federal Science Policy Office; the Fonds pour la Formation \`a la Recherche dans l'Industrie et dans l'Agriculture (FRIA-Belgium); the Agentschap voor Innovatie door Wetenschap en Technologie (IWT-Belgium); the F.R.S.-FNRS and FWO (Belgium) under the ``Excellence of Science - EOS" - be.h project n. 30820817; the Ministry of Education, Youth and Sports (MEYS) of the Czech Republic; the Lend\"ulet (``Momentum") Program and the J\'anos Bolyai Research Scholarship of the Hungarian Academy of Sciences, the New National Excellence Program \'UNKP, the NKFIA research grants 123842, 123959, 124845, 124850 and 125105 (Hungary); the Council of Science and Industrial Research, India; the HOMING PLUS program of the Foundation for Polish Science, cofinanced from European Union, Regional Development Fund, the Mobility Plus program of the Ministry of Science and Higher Education, the National Science Center (Poland), contracts Harmonia 2014/14/M/ST2/00428, Opus 2014/13/B/ST2/02543, 2014/15/B/ST2/03998, and 2015/19/B/ST2/02861, Sonata-bis 2012/07/E/ST2/01406; the National Priorities Research Program by Qatar National Research Fund; the Programa Estatal de Fomento de la Investigaci{\'o}n Cient{\'i}fica y T{\'e}cnica de Excelencia Mar\'{\i}a de Maeztu, grant MDM-2015-0509 and the Programa Severo Ochoa del Principado de Asturias; the Thalis and Aristeia programs cofinanced by EU-ESF and the Greek NSRF; the Rachadapisek Sompot Fund for Postdoctoral Fellowship, Chulalongkorn University and the Chulalongkorn Academic into Its 2nd Century Project Advancement Project (Thailand); the Welch Foundation, contract C-1845; and the Weston Havens Foundation (USA).

\end{acknowledgments}

\bibliography{auto_generated}

\cleardoublepage \appendix\section{The CMS Collaboration \label{app:collab}}\begin{sloppypar}\hyphenpenalty=5000\widowpenalty=500\clubpenalty=5000\vskip\cmsinstskip
\textbf{Yerevan Physics Institute, Yerevan, Armenia}\\*[0pt]
A.M.~Sirunyan, A.~Tumasyan
\vskip\cmsinstskip
\textbf{Institut f\"{u}r Hochenergiephysik, Wien, Austria}\\*[0pt]
W.~Adam, F.~Ambrogi, E.~Asilar, T.~Bergauer, J.~Brandstetter, M.~Dragicevic, J.~Er\"{o}, A.~Escalante~Del~Valle, M.~Flechl, R.~Fr\"{u}hwirth\cmsAuthorMark{1}, V.M.~Ghete, J.~Hrubec, M.~Jeitler\cmsAuthorMark{1}, N.~Krammer, I.~Kr\"{a}tschmer, D.~Liko, T.~Madlener, I.~Mikulec, N.~Rad, H.~Rohringer, J.~Schieck\cmsAuthorMark{1}, R.~Sch\"{o}fbeck, M.~Spanring, D.~Spitzbart, W.~Waltenberger, J.~Wittmann, C.-E.~Wulz\cmsAuthorMark{1}, M.~Zarucki
% ...
\vskip\cmsinstskip
% ...
1:  Also at Vienna University of Technology, Vienna, Austria\\
% ...
\end{sloppypar}
\end{document}